\documentstyle[12pt,aaspp4]{article}

\lefthead{Hirano et al.}
\righthead{SiO in the IRAS 16293-2422 outflow}

\begin{document}

\title{SiO EMISSION IN THE MULTI-LOBE OUTFLOW ASSOCIATED WITH 
       IRAS 16293$-$2422\altaffilmark{1}}

\author{Naomi Hirano\altaffilmark{2,3}, Hitomi Mikami \altaffilmark{4}, 
        Tomofumi Umemoto\altaffilmark{5}, Satoshi Yamamoto\altaffilmark{6},
        and Yoshiaki Taniguchi\altaffilmark{7}}

\altaffiltext{1}{Based on observations made at the Nobeyama Radio 
Observatory (NRO).
Nobeyama Radio Observatory is a branch of the National Astronomical 
Observatory of Japan, 
an inter-university research institute operated by the Ministry of 
Education, Science,
Sports, and  Culture, Japan.}

\altaffiltext{2}{Laboratory of Astronomy and Geophysics,
Hitotsubashi University, Kunitachi, Tokyo, 186-8601, Japan}

\altaffiltext{3}{Present address: Department of Astronomical Science,
The Graduate University for Advanced Studies,
National Astronomical Observatory, Mitaka, Tokyo, 181-8588, Japan}

\altaffiltext{4}{Nobeyama Radio Observatory, Minamimaki, Minamisaku, Nagano, 
384-1305, Japan}

\altaffiltext{5}{National Astronomical Observatory, Mitaka, Tokyo 181-8588, Japan}

\altaffiltext{6}{Department of Physics and Research Center for the Early Universe, 
The University of Tokyo, Bunkyo-ku, Tokyo 113-0033, Japan}

\altaffiltext{7}{Astronomical Institute, Graduate School of Science, Tohoku
     University, Aramaki, Aoba, Sendai 980-8578, Japan}

\begin{abstract}

We have mapped the thermal emission line of SiO ($v$ = 0; $J$ = 2--1) associated 
with the quadrupolar molecular outflow driven by the very cold far-infrared 
source IRAS 16293$-$2422.
The SiO emission is significantly enhanced in the northeastern red lobe
and at the position $\sim$50$''$ east of the IRAS source.
Strong SiO emission observed at $\sim$50$''$ east 
of the IRAS source presents evidence 
for a dynamical interaction between a part of the eastern blue lobe
 and the dense ambient gas condensation, however,
such an interaction is unlikely to be responsible for producing the 
quadrupolar morphology.
The SiO emission in the northeastern red lobe shows the spatial 
and velocity structure similar to those of the CO outflow, 
suggesting that the SiO emission comes from the molecular outflow 
in the northeastern red lobe itself.
The observed velocity structure is reproduced by a simple 
spatio-kinematic model of bow shock with a shock velocity of 
19--24 km s$^{-1}$ inclined by 30--45$^{\circ}$ from the plane of the sky.
This implies that the northeastern red lobe is independent of the 
eastern blue lobe and that the quadrupolar structure is due to two 
separate bipolar outflows.

The SiO emission observed in the western red lobe has a broad
pedestal shape with low intensity.
Unlike the SiO emission in the northeastern red lobe, the spatial
extent of the SiO emission in the western red lobe is restricted to 
its central region.
The spatial and velocity structures and the line profiles suggest
that three different types of the SiO emission are observed in this
outflow; the SiO emission arises from the interface
between the outflowing gas and the dense ambient gas clump,
the SiO emission coming from the outflow lobe itself, 
and the broad SiO emission with low intensity observed at the
central region of the outflow lobe.

\end{abstract}

\keywords{ISM: individual (IRAS 16293$-$2422) {\em -} ISM: jets and outflows {\em -} 
ISM: molecules {\em -} stars: formation {\em -} shock waves}

\section{INTRODUCTION}

Thermal emission from the SiO molecule is observed in shocked regions 
associated with molecular outflows from young stellar objects, 
in which the SiO abundance is enhanced by several orders of magnitude 
(e.g. \markcite{Bac91,Mik92,Mar92}Bachiller, Mart\'{\i}n-Pintado, \& Fuente 1991;
Mikami et al. 1992; Mart\'{\i}n-Pintado, Bachiller, \& Fuente 1992).
Such an abundance enhancement of SiO is interpreted as the release of Si or SiO 
from dust grains by means of strong shocks 
(e.g. \markcite{Mik92,Mar92}Mikami et al. 1992; Mart\'{\i}n-Pintado et al. 1992).
Since the SiO emission has been seldom detected 
in quiescent dark clouds because of its very low abundance
(of the order of 10$^{-12}$; \markcite{Ziu89,Mar92}Ziurys, Friberg,
\& Irvine 1989; Mart\'{\i}n-Pintado et al. 1992),
the SiO emission is generally believed to be 
a good tracer of shocked molecular gas.

Strong SiO emission observed in low-mass star forming regions
is mainly associated with
outflows driven by very cold far-infrared sources referred to as 
the ^^ ^^ Class 0$"$ sources (e.g., \markcite{Bac92, Bac96}Bachiller
\& G\'{o}mez-Gonz\'{a}lez 1992; Bachiller 1996).
On the basis of high-spatial resolution mappings of the SiO lines,
it has been revealed that both 
spatial and velocity structures of the shocked gas are closely related to 
the history and an ejection mechanism of the outflows. 
The representative cases are the outflows driven by the Class 0 sources
in the L1157 and L1448 dark
clouds; the SiO maps present a picture that these
 outflows are driven by  collimated and episodic jets ejected 
from the central star (\markcite{Zha95, Gue98, Dut97}Zhang et al. 1995;
Gueth, Guilloteau, \& Bachiller 1998; Dutrey, Guilloteau, \& Bachiller 1997).
Therefore, mapping of the SiO emission is expected to
provide an opportunity to study the origin of the molecular outflows
with complex morphology.

In this paper, we present the map of the SiO ($v$ = 0;$J$ = 2--1) 
emission in the quadrupolar outflow 
associated with
IRAS 16293$-$2422 (hereafter referred to as I16293).
I16293 is known to be a proto-binary system with a projected 
separation of 840 AU (\markcite{Mun92,Wal93}Mundy et al. 1992; 
Walker, Carlstrom, \& Bieging 1993) embedded in the nearby (160 pc) 
$\rho$ Ophiuchi molecular cloud.
This outflow consists of two pairs of lobes; the brighter one  
extending along the east-west direction and the other one 
along the northeast-southwest direction 
(\markcite{Wal99, Miz90}Walker et al. 1988; Mizuno et al. 1990).
To explain the conspicuous quadrupolar morphology of this outflow system,
\markcite{Wal93}Walker et al. (1993) proposed that the 
two pairs of lobes 
correspond to the two independent bipolar outflows driven 
by two independent sources.
On the other hand, \markcite{Miz90}Mizuno et al. (1990) showed  that the 
outflow is dynamically interacting with the dense ambient gas clump, and
suggested that a single outflow lobe could be split into
two lobes by the interaction.
To examine which idea is plausible, and to probe the
effect of the complex outflow on its surroundings, 
we investigate  the spatial and kinematic 
structure of the shocked gas traced by the SiO emission.

\section{OBSERVATIONS}

Observations of the SiO ($v$=0; $J$=2--1; 86.8 GHz) line were carried out 
with the Nobeyama 45 m telescope in 1992 April, 1993 April, and 1994 April.
We used the SIS receiver with a single-sideband (SSB) system temperature of $\sim$350 K.
The half-power beam width and the main beam efficiency were 20$''$ and 0.5, respectively.
The backend was a bank of eight high-resolution acousto-optical spectrometers 
with a frequency resolution of 37 kHz and a band width of 40 MHz,
which correspond to a velocity resolution of 0.13 km s$^{-1}$ 
and a velocity coverage of 140 km s$^{-1}$ at 86.8 GHz.
In addition to the SiO line, we also observed the H$^{13}$CO$^+$ ($J$=1--0) 
line simultaneously with the same receiver.
An area of 4.7$'$ $\times$ 3.0$'$ in Right Ascension and Declination was 
mapped with grid spacings of 14.1$''$ and 20$''$. 
The map center was taken to be the position of the southeastern millimeter 
continuum source, $\alpha$(1950) = 16$^{\rm h}$ 29$^{\rm m}$ 21.1$^{\rm s}$ 
and $\delta$(1950) = $-$24$^{\circ}$ 22$'$ 15.9$''$.
Offsets from this position are given as (${\Delta}{\alpha}$, ${\Delta}{\delta}$)
in units of arcsecond.
All observations were made in the position switching mode. 
Linear baselines were 
subtracted from the spectra.
A typical rms noise level was $\sim$0.16 K in $T_{\rm R}^{*}$.
The telescope pointing accuracy was checked by observing the nearby SiO maser 
source toward V446 Oph at 43 GHz and was better than 5$''$.

\section{RESULTS}

\subsection{Spatial Distribution and Kinematics of the Shocked Gas}

Figure 1 shows a map of the SiO emission integrated over $V_{\rm LSR}$ 
= 2 to 20 km s$^{-1}$ superposed on the CO $J$=1--0 outflow map 
presented by \markcite{Miz90}Mizuno et al. (1990).
The SiO $J$=2--1 emission is detected in this velocity range.
The SiO emission shows the U-shaped distribution open to the north with
 significant intensity enhancement in two clumps labeled E1 and E2 in Figure 1.
The northeastern peak E2 (+90, +40) coincides with 
the peak of the northeastern red (hereafter referred to as NER) lobe in the CO map.
On the other hand, the E1 peak (+50, 0) has no 
counterpart in the CO outflow map.
The SiO emission is also detected at the position of I16293; 
however, its intensity is rather weak.
At this position, detection of the higher transition lines of SiO
$J$ = 3--2, 5--4, 6--5, and 8--7
were reported by \markcite{Bla94}Blake et al. (1994) and
\markcite{Cec00}Ceccarelli et al. (2000).
The SiO emission observed toward the western red (WR) lobe shows
two local intensity maxima.
The position of one peak labeled W1 coincides with the peak of the
high-velocity CO emission in the WR lobe, while the other peak W2
is located at the northern boundary of the WR lobe.
No significant SiO emission is detected with $V_{\rm LSR}$ smaller 
than  2 km s$^{-1}$ in the whole mapped region.
Since most of the SiO emission is redshifted with respect to the cloud 
systemic velocity of $\simeq$ 4 km s$^{-1}$, 
the dense ambient gas interacting with the outflow is likely to 
exist behind the outflow lobes.

Figure 2 shows a set of velocity channel maps of the SiO $J$ = 2--1
emission from $V_{\rm LSR}$ = 2 to 20 km s$^{-1}$.
It is shown that the E1 component is seen in the velocity range from
$V_{\rm LSR}$ = 2 km s$^{-1}$ to 10 km s$^{-1}$,
while the E2 component contains higher velocity gas with maximum
velocity reaching $V_{\rm LSR}$ = 20 km s$^{-1}$.
Figures 2e--2i show that the peak position of the E2 component
in the velocity range from $V_{\rm LSR}$ = 10 km s$^{-1}$ to 
20 km s$^{-1}$ moves to the downstream as the velocity
increases.

The SiO line profiles observed at six representative positions,
which are the IRAS position, E1, upstream and downstream of 
E2, W1, and W2,
are plotted in Figure 3.
The SiO line at the IRAS position (Figure 3a) peaks at $V_{\rm LSR}$ = 
4.4 km s$^{-1}$ with its main-beam brightness temperature of 0.56 K.
The peak velocity of the SiO line is nearly the same as that of the 
H$^{13}$CO$^+$ line. 
However, the line width of the SiO emission (${\Delta}V_{\rm FWHM}$ = 3.5 km s$^{-1}$, where ${\Delta}V_{\rm FWHM}$ is the full width
of half maximum) 
is much broader than that of the H$^{13}$CO$^+$ emission 
(${\Delta}V_{\rm FWHM}$ = 1.1 km s$^{-1}$),  
indicating that the SiO emission arises from the perturbed gas component, 
and not from the quiescent gas traced by the H$^{13}$CO$^+$ line.
On the basis of analyses using $J$ = 5--4, 6--5, and 8--7
emission lines of SiO and its isotope, 
\markcite{Bla94}Blake et al. (1994) and \markcite{Cec00}
Ceccarelli et al. (2000) have revealed that the SiO emission 
observed at the IRAS position arises
from hot ($T >$60 K) and spatially compact ($\le$3$''$) region.
A typical line profile observed in the E1 clump (Figure 3b)
has a peak at the cloud 
systemic velocity of $V_{\rm LSR} \approx$ 4 km s$^{-1}$
with a gradual decrease in intensity toward the redshifted velocity
and a rapid decline toward the blueshifted velocity.
A line profile observed at the upstream of the E2 peak
(+80, +40; Figure 3c) shows a double peak structure
 at $V_{\rm LSR}$ = 5 and
12 km s$^{-1}$, while that at the downstream of the E2 peak
(+90, +70; Figure 3d) has its intensity maximum at 
$V_{\rm LSR} \sim$ 15 km s$^{-1}$.
At the position of W1 ($-$20, 0; Figure 3e), the SiO line is
pedestal shape without significant peak.
The SiO emission at W1 is seen from $V_{\rm LSR}$ = 2 km s$^{-1}$
to 13 km s$^{-1}$ with typical brightness temperatures of 0.2---0.4 K.
On the other hand, the SiO line observed near the W2 peak ($-$20, 60;
Figure 3f) has a peak at the ambient cloud velocity with rather narrow
line width of ${\Delta}V_{\rm FWHM} \sim$2 km s$^{-1}$.

\subsection{Comparison between the SiO and H$^{13}$CO$^+$
distributions}

In Figure 4, we compare the spatial distributions of the low-velocity 
($V_{\rm LSR}$ = 2 -- 8 km s$^{-1}$) and the high-velocity (V$_{\rm LSR}$ = 
8 -- 20 km s$^{-1}$) SiO emission  
with the quiescent dense gas traced by the H$^{13}$CO$^+$ emission 
($V_{\rm LSR}$ = 2 -- 6 km s$^{-1}$).
The H$^{13}$CO$^+$ map shows that there are two dense gas 
condensations in the mapped area.
One is associated with I16293 and its peak is at (+10, $-$10). 
Toward this condensation, the velocity dispersion of the 
H$^{13}$CO$^+$ emission ${\Delta}V_{\rm FWHM}$ is as large as $\sim$1.4 km s$^{-1}$,
which is the maximum value in the observed area.
Another one is located at $\sim 80^{\prime\prime}$ southeast of I16293. 
The H$^{13}$CO$^+$ line width in this condensation is 
$\sim$0.7--0.8 km s$^{-1}$, which is comparable to the typical line 
width of this molecule in the mapped region.
This southeastern condensation is also observed with the NH$_3$ (1, 1) line by 
\markcite{Miz90}Mizuno et al. (1990).
Hereafter we refer to this condensation as ^^ ^^ 16293E$"$.

Figure 4 shows that the spatial distribution of the SiO emission 
is anticorrelated with that of the H$^{13}$CO$^+$ emission;
especially, two prominent SiO clumps, E1 and E2, fall into 
the ^^ ^^ valley$"$ of the H$^{13}$CO$^+$ emission.
Since the spatial extent of the E2 clump is similar to that of the
NER lobe, the intensity drop of the H$^{13}$CO$^+$ toward the E2
clump suggests that the NER lobe is comprised of the ambient
gas swept-up by the wind blown out from the protostar.
It should be noted that the E1 clump that is dominated by the 
low-velocity emission is located 
between the two H$^{13}$CO$^+$ condensations.
Such a location of the E1 clump suggests that the SiO emission in this
clump arises from the region 
in which the outflow driven by I16293 is interacting with 16293E.
At the position of the E1 peak, the H$^{13}$CO$^+$ line width 
has the second largest value of ${\Delta}V_{\rm FWHM}$ = 1.1 km s$^{-1}$.
Such a line broadening supports the presence of dynamical interaction 
between the outflow and 16293E.
The interaction between the outflow and 16293E is discussed in 
the next section.

\section{DISCUSSION}

\subsection{Dynamical Interaction between the Outflow
and the Dense Gas Clump}

As mentioned in the previous section, the SiO emission in the E1
clump is likely to arise from the region where the outflow blown
out from the protostar impinge on the dense ambient
gas clump 16293E.
Since the line profile observed in the E1 clump that shows a 
strong cut-off 
at the ambient velocity and a gradual redshifted wing (Figure 3b),
it is suggested that the SiO emission in the E1 clump arises from the
quiescent material accelerated by the shock.
The dynamical interaction of the outflow with 16293E was also 
pointed out by \markcite{Miz90}Mizuno et al. (1990).
They suggested that the blueshifted gas blown eastwards 
from its driving source is interacting
with 16293E, and that such an interaction splits a single blueshifted 
lobe into the NER lobe and the eastern blue (EB) lobe, resulting in
the quadrupolar morphology.
However, the observed SiO emission is not blueshifted but redshifted;
this clearly contradicts the above scenario.

The redshifted SiO emission in the E1 clump suggests a possibility
that the redshifted gas in the NER lobe is 
blown out to the eastward from its driving source 
and changes its direction to the northeast by means of a dynamical 
interaction with 16293E.
In this case, the velocity field of the dense gas in 16293E
is expected to be disturbed by the momentum provided by the outflow.
On the basis of the CO observations, 
\markcite{Miz90}Mizuno et al. (1990) estimated the momentum of
the outflowing gas in the NER lobe to be 0.5 $M_{\odot}$ km s$^{-1}$.
Since the mass of 16293E is estimated to be $\sim$1.6 $M_{\odot}$
from the H$^{13}$CO$^+$ data by assuming the local thermodynamical
equilibrium (LTE) condition with the excitation temperature
$T_{\rm ex}$ =
15 K and optically-thin H$^{13}$CO$^+$ emission,
the gas of 16293E is expected to show 
the velocity shift of $\sim$0.3 km s$^{-1}$ toward the red from the
rest of the cloud.
However, there is no evidence of such a velocity shift in the 
H$^{13}$CO$^+$ data.
Therefore, we consider that the change of outflow direction by means
of a dynamical interaction is unlikely.
In addition, the velocity field observed in the E2 clump is explained by
the bow shock impinging on the ambient medium (see {\S}4.2),
suggesting that the outflowing gas in the NER lobe is blown out
to the northeast direction from its origin and does not change its
direction on its way to the E2 position.

Then, which outflow component is interacting with 16293E and
is responsible for the strong SiO emission at the E1 position?
A recent interferometric CO ($J$ = 1--0) map has shown
that the east-west pair of lobes, which corresponds to the EB and 
WR lobe pair, has an axis close to the plane of the sky and consists
of fan-shaped lobes, each of which contains both blueshifted and
redshifted components (Hirano et al. in preparation).
This implies that the EB lobe contains a considerable amount of 
redshifted gas in the far side of the lobe.
Therefore, we consider that the redshifted gas in the EB lobe is
interacting with 16293E and is producing the redshifted SiO emission
observed at E1.

\subsection{Simple Kinematic Model of the NER Lobe}

As shown in Figure 1, the spatial distribution of the SiO clump E2
resembles that of the NER lobe in the single-dish CO maps.
In particular, the map of the high-velocity SiO emission is quite similar to 
that of the high-velocity CO $J$ = 2--1 emission presented by 
\markcite{Cec98}Ceccarelli et al. (1998).
Channel maps presented in Figure 2 reveal that the high-velocity SiO
emission in the E2 clump increases its velocity downstream.
Such a velocity increase downstream is also observed
in the CO ($J$ = 2--1) emission in the NER lobe
(\markcite{Wal88}Walker et al. 1988).
These results indicate that
the high-velocity SiO emission in the E2 clump has both spatially and 
kinematically similar characteristics to the CO outflow in the 
NER lobe, suggesting that this SiO emission comes from the NER lobe itself.
Unlike the CO emission, the SiO emission in the E2 clump has another 
velocity component whose velocity is close to the cloud systemic velocity.
The presence of two velocity components is explained if the low- and the 
high-velocity SiO emission components come from the near side and the 
far side of the NER lobe, respectively.
Such two velocity components corresponding to the near sides and the far 
sides of the lobes were also seen in the position-velocity diagrams of the 
CO emission in the Mon R2 outflow (\markcite{Mey91}Meyers-Rice \& Lada 1991).
In the case of the NER lobe in the I16293 outflow, the low-velocity 
component has not been identified in the CO maps because 
it may be screened by overwhelming emission from the quiescent
 ambient gas.

In Figure 5, we show the position-velocity diagram of the SiO emission
along the major axis of the NER lobe (P.A. = 60$^{\circ}$).
The SiO emission in the NER lobe increases its 
velocity dispersion as an increasing distance from I16293, and
the largest velocity dispersion is observed toward 
slightly downstream of the E2 peak.
As pointed out by \markcite{Ben96}Bence, Richer, \& Padman (1996), 
such a velocity feature is explained if the SiO emission arises from the 
bow shock generated by the high-velocity jet impinging onto the ambient 
medium.
Recently, several authors have proposed the models that the molecular 
outflows are created through the propagation of the large bow shocks 
(e.g. \markcite{Rag93,Mas93}Raga \& Cabrit 1993; Masson \& Chernin 
1993).
In the following, we use a simple kinematical model of the bow shock 
proposed by \markcite{Har87}Hartigan, Raymond, \& Hartmann (1987) and 
compare the velocity features derived from the model with those 
observed.

The geometry of the bow shock is schematically depicted in Figure 6.
We assume that the bow shock has axial symmetry around the jet axis 
$z$ with a shape described by a function $z(r) = Ar^2 + Br^4$, 
where $r$ is the distance from the jet axis, and $A$ and $B$ are 
numerical constants.
In the frame of the exciting source, shock is propagating at 
a velocity of $V_{\rm s}$ into 
the ambient pre-shock medium moving at $V_{\rm amb}$.
In the frame of the bow shock, material impacts the shock surface with 
velocity of $V_{\rm s} - V_{\rm amb}$.
Since the velocity component normal to the surface of the bow shock 
$v_{\rm n} = (V_{\rm s} - V_{\rm amb}) {\rm cos}{\xi}$ diminishes
because of quick energy release via radiation, 
only the velocity component parallel to the surface 
$v_{\rm p} = (V_{\rm s} - V_{\rm amb}) {\rm sin}{\xi}$ is conserved.
Therefore, the post shock velocity is
$v_2 = (V_{\rm s} - V_{\rm amb}) {\rm sin} {\xi}$, 
and its angle is ${\theta} = ({\pi}/2) - {\xi}$.
The radial velocity $V_{\rm r}$ in the frame of the exciting source is given by
\begin{equation}
V_{\rm r} = (V_{\rm s} - V_{\rm amb}) {\rm sin}{\xi} {\rm sin} ({\xi} {\pm} {\phi}) - V_s {\rm cos} {\phi},
\end{equation}
where $\phi$ is the angle between the line of sight and the jet axis, 
and the + and $-$ signs refer to the top and bottom halves of the curve in 
Figure 6, respectively.
As demonstrated by \markcite{Har87}Hartigan et al. (1987), 
the total velocity dispersion $\Delta V$ of the bow-shock feature and 
the median velocity $V_{\rm median}$ are given by
$\Delta V = V_{\rm s} - V_{\rm amb}$
and
$V_{\rm median} = - [(V_{\rm s} + V_{\rm amb})/2] \times {\rm cos}{\phi}$, respectively.
In the case of the NER lobe, the total velocity dispersion is
 $\Delta V {\sim}$16 km s$^{-1}$, 
and $V_{\rm median}$ measured in the frame of the exciting source 
($V_{\rm LSR} \approx$4 km s$^{-1}$) is $\approx$8 km s$^{-1}$.
To estimate the inclination angle $\phi$, we compare the radial velocity 
distribution calculated from equation (1) with the observed 
position-velocity diagram.
If we assume that the bow shock has a shape described as 
$z(r) = 0.42 r^2 + 0.136 r^4$, which is designated as modified Raga shape 
in \markcite{Har87}Hartigan et al. (1987), the observed feature is 
reproduced with $\phi{\sim}135^{\circ}$ (i.e. the angle between the 
outflow axis and the plane of the sky {\it i} is $\sim$45$^{\circ}$), 
$V_{\rm s}$ = 19.3 km s$^{-1}$, and $V_{\rm amb}$ = 
3.3 km s$^{-1}$ (solid curve in Figure 5).
In the case of the bow shock with a parabolic shape, $z = r^2$, 
we obtained $\phi {\sim} 120^{\circ}$ ({\it i} $\sim$30$^{\circ}$), 
$V_{\rm s}$ = 24.0 km s$^{-1}$, and $V_{\rm amb}$ = 8.0 km s$^{-1}$ 
(dash-dotted curve in Figure 5).
Although the parameters depend on the assumed shape of the 
bow-shock surface, the observed velocity features can be explained if the 
axis of the flow is inclined by 30--45$^{\circ}$ from the plane of the sky 
and the shock velocity is $\sim$19--24 km s$^{-1}$.
The dynamical time scale of the E2 shock is estimated to be (5--7) $\times$ 10$^3$ yr.
It should be noted that the presence of the low-velocity component which 
is close to the cloud systemic velocity is a natural consequence in the 
context of the bow-shock outflow models.

As shown in Figure 5, the velocity pattern observed in the NER lobe is
reproduced by the simple bow-shock model, as in the case of the
SiO emission observed toward the red lobe of the bipolar outflow
in the L1448 dark cloud (\markcite{Dut97}Dutrey et al. 1997).
This implies that the NER lobe is one of the bipolar outflow lobes
and is not a splinter of the EB lobe.

The bow-shock model cannot explain the high-velocity emission of 
$V_{\rm LSR} {\sim}$18 km s$^{-1}$ observed farther downstream of the 
E2 peak (R.A. offset $>$ 120$''$).
Since the E2 peak is not located toward the end but toward the middle of 
the NER lobe, the bow shock at the E2 peak is unlikely to be a terminal 
shock.
Therefore,  
the NER lobe could have two-shock structure created by episodic ejection 
events as in the case of the blueshifted lobe of the L1157 outflow 
(e.g. \markcite{Zha95,Gue98}Zhang et al. 1995; Gueth et al. 1998).
To assess this possibility, higher-resolution observations with 
interferometer are required.

\subsection{Three Types of SiO Emission Observed in the I16293 
Outflow}

As discussed in the previous two subsections, the properties of the 
E1 and E2 clumps are different from each other.
The SiO emission in the E1 clump arises from the interface between the
outflow and the dense ambient gas clump, and that in the E2 clump
is likely to come from the outflow lobe itself.

The properties of the SiO emission observed at W2 appear to
be similar to those of the emission from E1, because the position
of W2 corresponds to the periphery of the WR lobe at which the 
outflowing gas faces the ambient medium.
The line profile of the SiO emission observed at W2 implies that
most of the shocked gas has the velocity close to the ambient velocity.

On the other hand, the SiO line profile observed at W1, where the
high-velocity CO emission is strongest in the 
WR lobe, shows a broad
pedestal shape with its width measured at 1$\sigma$ level is 
12 km s$^{-1}$.
Unlike the NER lobe, the CO emission in the WR lobe shows no systematic
displacement of emission peaks with increasing velocity, and has a peak
at the position close to that of W1 at all velocities
 (\markcite{Wal88}Walker et al. 1988).
The broad width
 of the SiO emission is likely to reflect the large velocity
dispersion of the outflowing gas along the line of sight 
toward the W1 position.
Unlike the E2 clump, the spatial extent of the SiO emitting region in
the W1 is restricted to the small area of $\sim$20$''$ in size,
which is much smaller than the size of the WR lobe observed in the
CO.
Such a broad SiO emission with low intensity is also observed
in the central region of the outflows in S140 and IC1396N
(Hirano et al. in preparation).

To summarize, in the I16293 outflow, the SiO emission lines having
three different properties are observed.
The first one is the enhanced SiO emission at the 
interface between the outflowing gas and the dense ambient gas
clumps (E1 and W2), which shows the line profile peaks at the
velocity close to the cloud systemic velocity.
The second one is the SiO emission showing similar spatial extent and 
kinematic structure as the CO outflow, and considered to arise from the
 outflow lobe itself (E2). 
The third one is the broad SiO emission with low intensity
observed toward the central region of the outflow lobe (W1).

\subsection{SiO Abundances}

To estimate the SiO abundance in the shocked gas, 
we assumed optically thin emission,
 and calculated the column densities of the SiO molecules by assuming
the LTE condition.
The permanent dipole moment of the SiO molecule is employed to be 3.1 Debye.
In the following, we assumed the excitation temperature $T_{\rm ex}$ 
to be 15 K, 
which is almost the same as the rotational temperature estimated from 
the NH$_3$ (1, 1) and (2, 2) transitions (\markcite{Miz90}Mizuno et al. 1990).
It is possible that the excitation temperature of the SiO in the shocked
gas is higher than that assumed here, because recent observations of 
high-excitation ammonia have revealed that the shocked gas in several
outflows is heated to more than 100 K
(e.g. \markcite{Bac93,Ume99}Bachiller, Mart\'{\i}n-Pintado, \& Fuente 
1993; Umemoto et al. 1999).
If we adopt 100 K as an excitation temperature, the column density 
of the SiO becomes a factor of three higher than that calculated 
for $T_{\rm ex}$ = 15 K.

By using the total integrated intensities (from $V_{\rm LSR}$ = 2 to 20 km s$^{-1}$), 
the SiO column densities, $N$(SiO), at the E1 and E2 positions
 are derived 
to be 4.3$\times$10$^{13}$ cm$^{-2}$ and
 8.4$\times$10$^{13}$ cm$^{-2}$, respectively, whereas
it is estimated to be $\simeq  2.0\times 10 ^{13}$ cm$^{-2}$
at both the W1 and W2 positions.
At the IRAS position, we obtain  $N$(SiO) to be $1.0\times 10 ^{13}$ cm$^{-2}$.
When we adopt an H$_2$ column density averaged over the core of 0.2 pc in radius, 
3.6$\times$10$^{22}$ cm$^{-2}$, derived from the C$^{18}$O observations 
of \markcite{Miz90}Mizuno et al. (1990),
the SiO abundance $X$(SiO) is estimated  to be 
1.2$\times$10$^{-9}$ toward the E1 peak, 5.2$\times$10$^{-10}$
toward W1 and W2, and 2.8$\times$10$^{-10}$ at the IRAS position.
It should be noted that the abundance values estimated here are the 
beam-averaged values.
If the SiO emission observed at the IRAS position comes from the 
small area of $<$3$''$ size, the $X$(SiO) becomes a factor of $\sim$50
higher than that estimated here, which is consistent with the SiO
abundance value derived by \markcite{Cec00}Ceccarelli et al. (2000)
from their LVG analysis.

To estimate the SiO abundance in the outflowing gas in the NER lobe,
we compare the  column density of the high-velocity SiO (from 8 to 20 km s$^{-1}$ )
with that of H$_2$ derived from the CO data observed by \markcite{Miz90}Mizuno et al. (1990).
At the E2 peak, the $N$(SiO) and $N$(H$_2$) of the high-velocity gas are 
estimated to be 7.3$\times$10$^{13}$ cm$^{-2}$ and 
8.8$\times$10$^{20}$ cm$^{-2}$, respectively.
We thus obtain $X$(SiO) in the high-velocity gas to 
be 8.3$\times$10$^{-8}$ at the E2 peak.
This indicates that the SiO abundance in the NER lobe is comparable 
to that observed in the blueshifted lobe of the L1157 outflow 
(\markcite{Mik92}Mikami et al. 1992) 
and is enhanced by a factor of $\sim$10$^{4}$ as compared to the upper 
limits in the dark clouds (\markcite{Zir89}Ziurys et al. 1989).
If we assume that all the Si atoms 
(Si/H$_2$ $\sim$7$\times$10$^{-5}$) is depleted 
on grains in the unperturbed component, 
at least $\sim$0.1 \% of Si in the
interacting region should be released in the gas phase in order to
explain the SiO abundance in the NER lobe.

\section{Summary}
We have mapped the thermal emission line of SiO ($J$ = 2--1) associated
with the quadrupolar molecular outflow driven by I16293.
Our main results are the following:

1. The SiO emission is significantly enhanced in the northeastern red
lobe (E2) and at the position $\sim$50$''$ east of the IRAS source (E1).
The SiO emission is also detected at the IRAS position and toward the
western red lobe (W1 and W2),
however, the line intensities observed
at these positions are less than half of those observed at E1 and E2.

2. Most of the SiO emission is redshifted with respect to the cloud
systemic velocity of $V_{\rm LSR}$ = 4 km s$^{-1}$.
This suggests that the dense ambient gas interacting with the outflow
exists behind the outflow lobes.

3. The E1 clump is dominated by the low-velocity emission 
($V_{\rm LSR}$ = 2 -- 10 km s$^{-1}$) and is located between the dense 
gas condensation associated with I16293 and the other one 16293E,
which is located at $\sim$80$''$ southeast of I16293.
The location of the E1 clump and the SiO line profile observed there
suggest that
the SiO emission in the E1 clump arises from the region
where the outflow blown out from I16293 is dynamically interacting
with 16293E.

4. In the E2 clump, high velocity SiO emission with maximum velocity 
reaching $V_{\rm LSR}$ = 20 km s$^{-1}$ is observed.
The spatial and velocity structure of the SiO emission in the E2
clump are similar to those of the NER lobe observed in the CO emission, 
suggesting that the
SiO emission in the E2 clump comes from the NER lobe itself.
The velocity pattern of the SiO in the NER lobe is reproduced with a 
simple spatio-kinematic model of bow shock with a shock velocity
of 19--24 km s$^{-1}$ inclined by 40--45$^{\circ}$ from the plane 
of the sky.
This model gives a dynamical timescale of $\sim$5000--7000 yr
for the NER lobe, being consistent with the significant increase of
the SiO abundance by a factor of $\sim$10$^4$ relative to the
values in dark clouds.

5. Although the SiO emission at E1 presents evidence for a
dynamical interaction between a part of the EB lobe and
16293E, such an interaction is unlikely to be responsible for producing
the quadrupolar morphology.
Since the velocity structure of the NER lobe, which is reproduced by the 
bow-shock model, is similar to that of the lobe in the bipolar
outflow,
the NER lobe is likely to be one of the lobes of the bipolar
outflow rather than a splinter of the EB lobe.
Therefore, we consider that contribution of two independent 
bipolar outflows is the
plausible explanation for the origin of the quadrupolar structure of the 
I16293 outflow.

6. In the I16293 outflow, the SiO emission lines with three different 
properties are observed. They are 1) the enhanced SiO emission at
the interface between the outflow and dense ambient gas (E1 and W2),
2) the SiO emission arises from the outflow lobe itself (E2), 
and 3) the broad SiO emission with low intensity observed toward the
central region of the outflow lobe (W1).

\acknowledgments

We would like to thank the staff of NRO for the operation of our observations 
and their support in data reduction. 
N.H. acknowledges support from a Grant-in-Aid from the Ministry of Education, 
Science, Sports, and Culture (No. 09640315).

\clearpage

\figcaption[fig1.eps]{Integrated intensity map of the SiO $J$ = 2--1 emission 
($V_{\rm LSR}$ = 2--20 km s$^{-1}$; {\it gray scale}) 
superposed on the CO $J$ = 1--0 outflow map ($V_{\rm LSR} = -$10--2 km s$^{-1}$; 
{\it blue contours} and  $V_{\rm LSR}$ = 6--20 km s$^{-1}$; {\it red contours}) 
presented by Mizuno et al. (1990). 
The contours of the SiO $J$ = 2--1 map are drawn every 1.0 K km s$^{-1}$ (3 $\sigma$) 
with the lowest contours of 1.0 K km s$^{-1}$ (3 $\sigma$).
Two crosses indicate the positions of the I16293 binary protostars.
Green line denotes the cut of the position-velocity diagram presented in Figure 5.
\label{fig1}}

\figcaption[fig2.eps]{Channel-velocity maps of the SiO $J$ = 2--1
emission with the velocity interval of 2 km s$^{-1}$.
Contours are every 0.33 K km s$^{-1}$ (3 $\sigma$) with the 
lowest contours of 0.33 K km s$^{-1}$ (3 $\sigma$).
Solid and dashed grey lines denote the areas of the red and blue
lobes of the CO $J$ = 1--0 outflow (Mizuno et al. 1990), respectively.
The positions of two protostars are indicated with crosses.
\label{fig2}}

\figcaption[fig3.eps]{Line profiles of the SiO $J$ = 2--1 emission observed toward 
(a) I16293, (b) E1, (c) the upstream of the E2 peak,
(d) the downstream of the E2 peak, e) W1, and f) W2.
The offsets are in arcseconds from the position of the southeastern continuum source: 
$\alpha$(1950) = 16$^{\rm h}$ 29$^{\rm m}$ 21.$^{\rm s}$1, 
$\delta$(1950) = $-24^{\circ}$ 22$'$ 15.9$''$.
\label{fig3}}

\figcaption[fig4.eps]{Comparison of the high-velocity SiO ($V_{\rm LSR}$ = 8--20 km s$^{-1}$; 
{\it red contours}), 
the low-velocity SiO ($V_{\rm LSR}$ = 2--8 km s$^{-1}$; {\it green contours}), 
and the H$^{13}$CO$^+$ $J$ = 1--0 integrated intensity ($V_{\rm LSR}$ = 
2--6 km s$^{-1}$; {\it gray scale}) maps.
The contours for the high-velocity SiO are drawn every 0.69 K km s$^{-1}$ (3 $\sigma$) 
from 0.69 K km s$^{-1}$ (3 $\sigma$), 
and those for the low-velocity SiO are every 0.60 K km s$^{-1}$ (3 $\sigma$) 
from 0.60 K km s$^{-1}$ (3 $\sigma$).
The contours of the H$^{13}$CO$^+$ map are drawn every 0.54 K km s$^{-1}$ (3 $\sigma$) 
with the lowest contours of 0.54 K km s$^{-1}$ (3 $\sigma$).
The crosses denote the positions of the protostars. 
\label{fig4}}

\figcaption[fig5.eps]{Position-velocity diagram of the SiO $J$ = 2--1 emission 
along the major axis of the NER lobe (PA = 60$^{\circ}$; green line in Figure 1).
The lowest contour level and the contour spacing are 0.3 K.
The solid and dash-dotted curves represent the velocity patterns
derived from the bow-shock outflow models with modified Raga shape 
and parabolic shape, respectively.
\label{fig5}}

\figcaption[fig6.eps]{Geometry of the bow shock 
impacting a downstream medium moving at $V_{\rm amb}$ in the frame of the exciting source.
The shape of the bow shock is given by $z(r)$.
$\phi$ is the inclination angle between the axis of the bow shock and 
the line of sight ($\phi = \pi / 2 + i$).
The shock propagation velocity is $V_{\rm s}$.
\label{fig6}}

\end{document}